# Photonic-enabled radio-frequency self-interference cancellation incorporated in an in-band full-duplex radio-over-fiber system

Taixia Shi[a,b], Yu Chen[a,b] and Yang Chen[a,b,*]

[a] Shanghai Key Laboratory of Multidimensional Information Processing, East China Normal University, Shanghai 200241, China
[b] Engineering Center of SHMEC for Space Information and GNSS, East China Normal University, Shanghai 200241, China
[*] ychen@ce.ecnu.edu.cn

**ABSTRACT**
A photonic approach for radio-frequency (RF) self-interference cancellation (SIC) incorporated in an in-band full-duplex radio-over-fiber system is proposed. A dual-polarization binary phase-shift keying modulator is used for dual-polarization multiplexing at the central office (CO). A local oscillator signal and an intermediate-frequency signal carrying the downlink data are single-sideband modulated on the two polarization directions of the modulator, respectively. The optical signal is then transmitted to the remote unit, where the optical signals in the two polarization directions are split into two parts. One part is detected to generate the up-converted downlink RF signal, and the other part is re-modulated by the uplink RF signal and the self-interference, which is then transmitted back to the CO for the signal down-conversion and SIC via the optical domain signal adjustment and balanced detection. The functions of SIC, frequency up-conversion, down-conversion, and fiber transmission with dispersion immunity are all incorporated in the system. An experiment is performed. Cancellation depths of more than 39 dB for the single-tone signal and more than 20 dB for the 20-MBaud 16 quadrature amplitude modulation signal are achieved in the back-to-back case. The performance of the system does not have a significant decline when a section of 4.1-km optical fiber is incorporated.

## 1. Introduction

In recent years, with the rapid development of mobile communication, artificial intelligence, Internet of Things, big data and other industries, the number of wireless communication users and networking devices has increased dramatically. Various communication devices have higher requirements for communication traffic and speed. However, the spectrum resources are limited, so it is necessary to make reasonable and full use of the spectrum. The in-band full-duplex (IBFD) system can double the spectrum efficiency by employing the same frequency in the downlink and uplink [1, 2], which is a promising solution to the limited spectrum resources. Radio-over-fiber (ROF) [3] system directly transmits the RF analog signal in the optical fiber, which greatly simplifies the remote unit (RU) and is considered to be a promising solution for the further wireless communication systems. ROF systems can also be operated in the IBFD condition. However, the

IBFD operation introduces a strong self-interference from the transmitting antenna to the receiving antenna, and the self-interference cannot be filtered out because of the identical frequency in the downlink and uplink. Therefore, it is very important to eliminate the self-interference effectively.

In [4], Duarte et al. demonstrated the IBFD transmission of 2.4-GHz WiFi signals through experiments. Since then, many studies for self-interference cancellation (SIC) are conducted. Because the power of the self-interference is much greater than that of the signal of interest (SOI), different kinds of methods should be combined to achieve a large cancellation depth for IBFD systems. The SIC can be implemented in the antenna domain [5], the analog domain [6], and the digital domain[7, 8]. The digital domain method is a low-cost solution to SIC. However, without the SIC in the antenna domain and the analog domain, the power of the self-interference will exceed the dynamic range of the analog to digital converter. Thus, many works have been conducted for analog SIC. Besides implementing the SIC using conventional electronic circuits, the SIC can also be realized based on microwave photonics[9, 10], which has the unique advantages of large bandwidth, high frequency, good tunability, and immunity to electromagnetic interference. Furthermore, the photonic-based method can be seamlessly combined with the ROF systems.

Some photonic-based SIC methods were proposed in the past decade. In [11], a method using two independent laser sources and two parallel Mach-Zehnder modulators (MZMs) biased at adjacent quadrature transmission points was proposed for SIC. To simplify the system structure, some simplified architectures were proposed. In [12], two electro-absorption modulated lasers and a balanced photodiode (BPD) were used in conjunction to realize the cancellation of the self-interference. A single laser, a single modulator and a single photodetector (PD) were cascaded for SIC in [13, 14], which simplifies the photonic-based SIC system structure greatly and avoids the instability caused by using two independent optical paths. However, the SIC methods mentioned above only focuses on the function of SIC.

In practical applications, the elimination of the self-interference needs to be combined with the system structure. Therefore, some photonic-based SIC methods were designed in combination with the characteristics of the IBFD ROF system. Since the IBFD ROF system with long-distance fiber transmission will be influenced by the fiber dispersion, some methods were proposed in [15-17] to overcome the influence on the performance of SIC and the distortion on the SOI caused by the fiber dispersion. Furthermore, for an IBFD ROF system, it is highly desirable that the SIC can be implemented in conjunction with some other key functions of the system. Therefore, some methods that combine the function of SIC and frequency down-conversion were proposed in [15, 17, 18]. In [19], the SIC in an IBFD ROF system is further combined with the entire ROF system structure. In the system, the self-interference is transmitted back to the central office (CO), where the SIC is carried out. By moving the SIC from the RU to the CO, the complexity of the CO can be reduced.

However, all the photonic-based SIC methods mentioned above cannot simultaneous realize all the functions, i.e., overcoming the influence on the SIC and the distortion of the SOI caused by the fiber dispersion, realizing the SIC in conjunction with frequency conversion, and moving the SIC from the RU to CO to simplify the RU. To solve this problem, in this paper, a photonic approach for SIC incorporated in an IBFD ROF system is proposed. In the system, a dual-polarization binary phase-shift keying (DP-BPSK) modulator is used for polarization multiplexing at the CO. A local oscillate (LO) signal and an intermediate-frequency (IF) signal carrying the downlink data are single-sideband modulated on the two polarization directions of the modulator, respectively. The optical signal is then transmitted to the RU, where the optical signals in the two polarization directions are split into two parts. One part is detected to generate the up-converted downlink RF signal, and the other part is re-modulated by the uplink signal and the self-interference, which is then transmitted back to the CO for the signal down-conversion and SIC via optical domain signal adjustment and balanced detection. An experiment is performed. Cancellation depths of more than 39 dB for the single-tone signal and more than 20 dB for the 20-MBaud 16 quadrature amplitude modulation (16-QAM) signal are achieved in the back-to-back case. The performance of the system does not have a significant decline when a section of 4.1-km optical fiber is incorporated.

## 2. Principle

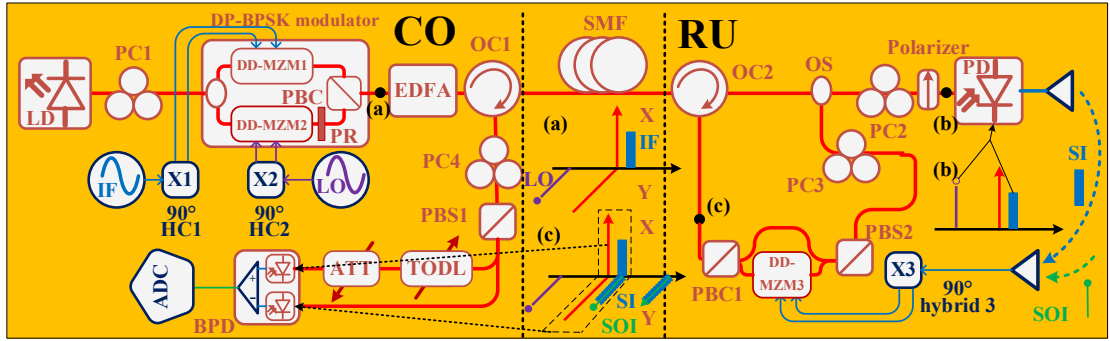

Fig. 1. Schematic diagram of the proposed photonic-based SIC system incorporated in an IBFD ROF system. CO, central office; RU, remote unit; LD, laser diode; PC, polarization controller; HC, hybrid coupler; DP-BPSK modulator, dual-polarization binary phase-shift keying modulator; DD-MZM, dual-drive Mach-Zehnder modulator; PR, polarization rotator; PBC, polarization beam combiner; OC, optical circulator; SMF, single-mode fiber; EDFA, erbium-doped fiber amplifier; OS, optical splitter; PBS, polarization beam splitter; PD, photodetector; TODL, tunable optical delay line; ATT, attenuator; BPD, balanced photodetector; ADC, analog to digital converter; IF, intermediate frequency; LO, local oscillator; SI, self-interference; SOI, signal of interest.

Figure 1 shows the schematic diagram of the proposed photonic-based SIC system incorporated in an IBFD ROF system. A continuous-wave (CW) light wave from a laser diode (LD) is injected into a DP-BPSK modulator at the CO. The DP-BPSK modulator is composed of a 3-dB optical splitter, two dual-drive Mach-Zehnder modulators (DD-MZMs), a 90° polarization rotator, and a polarization beam combiner (PBC). In the system, the two DD-MZMs are both biased as single-sideband (SSB) modulators. The two RF ports of the X-polarization DD-MZM1 are driven by an IF signal carrying the downlink data via a 90° hybrid coupler (HC1), whereas those of the Y-polarization DD-MZM2 are driven by an LO signal via a second 90° hybrid coupler (HC2). The schematic of the optical spectrum from the DP-BPSK modulator is shown in Fig. 1(a). Note that the 1st-order optical sidebands generated by the LO signal and the IF signal are in the opposite directions. Then, the optical signal is amplified by an erbium-doped fiber amplifier (EDFA) before transmitting to the RU via a section of single-mode fiber (SMF). An optical circulator (OC1) is inserted between the EDFA and the SMF for separating the uplink optical signal from the RU.

After the downlink fiber transmission, the downlink optical signal is received at the RU, which is firstly sent to OC2, and then split into two parts by an optical splitter (OS). One part of the optical signal is sent to a PD via PC2 and a polarizer. By properly tuning PC2, the optical signals in the two polarization directions are combined, which is shown in Fig. 1 (b). By beating the combined optical signal in the PD, the original IF signal can be up-converted to an RF signal, which can be further filtered and then radiated from the transmitting antenna as the downlink RF signal. The other part of the optical signal from the OS is sent to a polarization beam splitter (PBS2) via PC3 to separate the optical signals at the two orthogonal polarization directions. The optical signal in the Y polarization is sent to a DD-MZM (DD-MZM3), where it is also SSB-modulated by the received signal at the RU, consisting of the uplink RF signal and the self-interference from the transmitting antenna. The modulated optical signal in the Y polarization from DD-MZM3 and the optical signal in the X polarization is combined in PBC1 and transmitted back to the CO via OC2 and the SMF. The schematic of the uplink optical signal is shown in Fig. 1 (c).

At the CO, the uplink optical signal is sent to PBS1 via OC1 and PC4, where the optical signals at the two orthogonal polarization directions are separated and sent to a BPD. When only the optical sideband close to the optical carrier as shown in the dotted box in Fig. 1(c) is considered, the beating product of the X polarization can be used as an IF reference, and that of the Y polarization consists not only a down-converted uplink IF signal but also a self-interference. To cancel the self-interference, an optical attenuator (ATT) and a tunable optical delay line (TODL) are set on one optical path to the BPD to match the delay and amplitude of the reference and the self-interference.

Assuming the IF signal and the LO signal are

$$V_1(t) = A_1 \cos(\omega_{IF} t + \varphi_1), \tag{1}$$

$$V_2(t) = A_2 \cos(\omega_{LO} t + \varphi_2), \tag{2}$$

where $A_1$ and $A_2$ are the amplitudes, $\omega_{IF}$ and $\omega_{LO}$ are the angular frequencies, and $\varphi_1$ and $\varphi_2$ are the phases of the IF signal and the LO signal, respectively.

Since the IF signal and the LO signal are applied to the X-polarization DD-MZM1 and Y-polarization DD-MZM2 of the DP-BPSK modulator, respectively, the optical signal from the DP-BPSK modulator can be expressed as

$$\mathbf{E}_{DP-BPSK}(t) = \begin{bmatrix} E_X(t) \\ E_Y(t) \end{bmatrix} \\ \propto \begin{bmatrix} \frac{\sqrt{2}}{2} J_0(m_1) \exp\left(j\omega_c t + j\frac{1}{4}\pi\right) + J_1(m_1) \exp(j\omega_c t - j\omega_{IF} t - j\varphi_1 + j\pi) \\ \frac{\sqrt{2}}{2} J_0(m_2) \exp\left(j\omega_c t + j\frac{1}{4}\pi\right) + J_1(m_2) \exp(j\omega_c t + j\omega_{LO} t + j\varphi_2 + j\pi) \end{bmatrix}, \tag{3}$$

where $E_X(t)$ and $E_Y(t)$ represent the optical signals of the two DD-MZMs, $m_i = \pi V_i / V_\pi \ (i=1,2)$ is the modulation index, and $\omega_c$ is the angular frequency of the input optical signal. In Eq. (3), only the optical carrier and the 1st-order optical sideband are considered.

The polarization multiplexed optical signal is transmitted to the RU by a section of SMF. It is noted that due to the single-sideband operation, the system is immune to the power fading effect caused by the fiber dispersion. Then, the optical signal is split into two parts at the RU, which are respectively used for the generation of the downlink RF signal and the cancellation of the self-interference. The former part is applied to a polarizer via PC2 with its principal axis oriented at an angle of 45° to one principle axis of the DP-BPSK modulator, and then detected by the PD to obtain the downlink signal. The signal in the RF band that we are interested in can be expressed as

$$V_{Trans}(t) \propto J_1(m_1) J_1(m_2) \cos(\omega_{IF} t + \omega_{LO} t + \varphi_1 + \varphi_2). \tag{4}$$

Because of the crosstalk from the transmitting antenna to the receiving antenna, the self-interference at the receiving antenna can be expressed as

$$V_{SI}(t) = A_{SI} \cos\left[\omega_s(t-\tau_1) + \varphi_1 + \varphi_2\right], \tag{5}$$

where $A_{SI}$ is the amplitude of the self-interference, $\tau_1$ is the time delay of the self-interference, and $\omega_s = \omega_{IF} + \omega_{LO}$, representing the angular frequency of the self-interference and the SOI.

For the convenience of analysis, the SOI is ignored and only the self-interference in the received signal is considered. The self-interference is applied to DD-MZM3, which also functions as an SSB modulator. The optical signal from DD-MZM3 can be expressed as

$$\begin{aligned}
E_{Y-SSB}(t) \propto & \frac{1}{2} J_0(m_2) J_0(m_3) \exp\left(j\omega_c t + j\frac{1}{2}\pi\right) \\
& + \frac{\sqrt{2}}{2} J_1(m_2) J_0(m_3) \exp\left(j\omega_c t + j\omega_{LO} t + j\varphi_2 + j\frac{5}{4}\pi\right) \\
& + \frac{\sqrt{2}}{2} J_0(m_2) J_1(m_3) \exp\left(j\omega_c t - j\omega_s t + j\omega_s \tau_1 + j\frac{5}{4}\pi - j\varphi_1 - j\varphi_2\right) \\
& + J_1(m_2) J_1(m_3) \exp\left(j\omega_c t + j\omega_{LO} t - j\omega_s t + j\omega_s \tau_1 - j\varphi_1\right)
\end{aligned} \tag{6}$$

where $m_3 = \pi A_{SI} / V_\pi$.

Then the optical signal in the X polarization and the modulated optical signal in the Y polarization are transmitted back to the CO. At the CO, the optical signals in the two polarizations are separated by PBS1, which are injected into the two optical ports of the BPD, respectively. Between PBS1 and the BPD, an ATT and a TODL are employed to adjust the amplitude and the delay of the optical signal in X polarization while the optical signal in Y polarization is unchanged. The two optical signals applied to the BPD can be expressed by

$$\begin{aligned}
E_{BPD-X}(t) &\propto \sqrt{\alpha} E_X(t - \tau_2) \\
&= \frac{\sqrt{2}}{2} \sqrt{\alpha} J_0(m_1) \exp\left(j\omega_c(t - \tau_2) + j\frac{1}{4}\pi\right) \\
&+ \sqrt{\alpha} J_1(m_1) \exp\left(j\omega_c(t - \tau_2) - j\omega_{IF}(t - \tau_2) - j\varphi_1 + j\pi\right)
\end{aligned} \tag{7}$$

$$E_{BPD-Y}(t) = E_{Y-SSB}(t), \tag{8}$$

where $\alpha$ is the power attenuation coefficient introduced by the ATT and $\tau_2$ is the delay introduced by the TODL.

The photocurrent in the IF band form the BPD is

$$\begin{aligned}
i(t) &= i_X(t) - i_Y(t) \\
&\propto \sqrt{2}\alpha J_0(m_1) J_1(m_1) \cos\left(\omega_{IF} t - \omega_{IF} \tau_2 + \varphi_1 - \frac{3}{4}\pi\right) \\
&- J_0(m_2) J_0(m_3) J_1(m_2) J_1(m_3) \cos\left(\omega_{IF} t - \omega_s \tau_1 + \varphi_1 + \frac{1}{2}\pi\right)
\end{aligned} \tag{9}$$

In order to cancel the self-interference, the amplitudes and the phases of the two terms in Eq. (9) should be matched, i.e., the following condition should be satisfied

$$\alpha=\frac{\sqrt{2}J_0(m_2)J_0(m_3)J_1(m_2)J_1(m_3)}{2J_0(m_1)J_1(m_1)}, \quad (10)$$

$$\omega_{IF}\tau_2=\omega_s\tau_1-\frac{5}{4}\pi. \quad (11)$$

It can be seen that there is a constant phase of -5π/4 in Eq. (11), which must be compensated by a phase shifter if the system is a wideband system. The phase compensation can be done immediately after the uplink signal is received at the RU. If the -5π/4 phase shift is compensated, the SIC can be achieved for a wideband system when $\tau_2=\omega_s\tau_1/\omega_{IF}$ establishes. For narrowband systems, in which the carrier frequency is far greater than the signal bandwidth, the phase compensation can be also be implemented by using the TODL. In this case, the time delay of the TODL should be

$$\tau_2=\frac{\omega_s\tau_1}{\omega_{IF}}-\frac{5}{4\omega_{IF}}\pi. \quad (12)$$

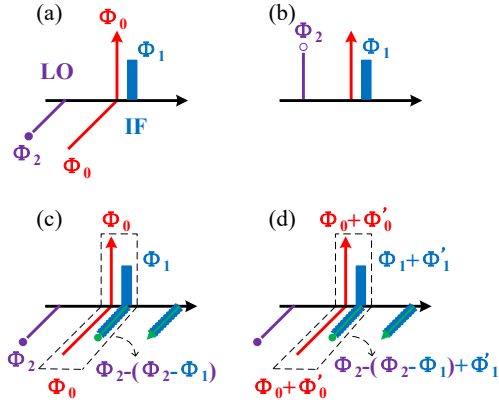

Fig. 2. The phase shifts introduced to the optical signals by the fiber dispersion at different locations in the system

In addition, the influence of the fiber dispersion on the SIC capability of the system is also analyzed. In the downlink, different phase shift is introduced to different optical wavelengths of the transmitted signal due to the fiber dispersion, as shown in Fig. 2(a). Here, we only consider the influence from the fiber dispersion, so only the additional phases those are not considered in the deduction above are shown in the figure. The optical signal received at the RU can be written as

$$\mathbf{E}_{downlink}(t)=\begin{bmatrix}E'_X(t)\\E'_Y(t)\end{bmatrix}$$
$$\propto \begin{bmatrix}\left\{\frac{\sqrt{2}}{2}J_0(m_1)\exp\left(j\omega_c t+j\frac{1}{4}\pi+\Phi_0\right)\\+J_1(m_1)\exp\left(j\omega_c t-j\omega_{IF}t-j\varphi_1+j\pi+\Phi_1\right)\right\}\\ \left\{\frac{\sqrt{2}}{2}J_0(m_2)\exp\left(j\omega_c t+j\frac{1}{4}\pi+\Phi_0\right)\\+J_1(m_2)\exp\left(j\omega_c t+j\omega_{LO}t+j\varphi_2+j\pi+\Phi_2\right)\right\}\end{bmatrix}, \quad (13)$$

where $\Phi_0$, $\Phi_1$ and $\Phi_2$ are the phase shifts introduced by the fiber dispersion to the optical carrier, the 1st-order optical sideband of the IF signal, and the opposite 1st-order optical sideband of the LO signal, respectively.

When the phase shifts introduced by the fiber dispersion is considered, an additional phase shift of $\Phi_2$-$\Phi_1$ is introduced to the downlink RF signal, which is also the self-interference and can be rewritten as

$$V_{SI-fiber}(t) = A_{SI}\cos\left[\omega_s(t-\tau_1)+\varphi_1+\varphi_2+\Phi_2-\Phi_1\right]. \qquad (14)$$

In the uplink, optical signal is SSB modulated by the self-interference at DD-MZM3, obtaining two optical sidebands as shown in Fig. 2(c). The optical carriers and the optical sidebands we are concerned about in the two dash boxes can be expressed as

$$\mathbf{E}_{uplink}(t) = \begin{bmatrix} E_X''(t) \\ E_Y''(t) \end{bmatrix}$$

$$\propto \begin{bmatrix} \left\{ \frac{\sqrt{2}}{2}J_0(m_1)\exp\left(j\omega_c t + j\frac{1}{4}\pi + \Phi_0\right) \\ +J_1(m_1)\exp\left(j\omega_c t - j\omega_{IF} t - j\varphi_1 + j\pi + \Phi_1\right) \right\} \\ \left\{ \frac{1}{2}J_0(m_2)J_0(m_3)\exp\left(j\omega_c t + j\frac{1}{2}\pi + \Phi_0\right) \\ +J_1(m_2)J_1(m_3)\exp\left(j\omega_c t + j\omega_{LO} t - j\omega_s t + j\omega_s \tau_1 - j\varphi_1 + \Phi_1\right) \right\} \end{bmatrix}. \qquad (15)$$

Then the optical signal is sent back to the CO via the SMF. Additional phase shifts of $\Phi_0'$ and $\Phi_1'$ are introduced to the optical carriers and the IF optical sidebands. Since the optical reference signal and the optical self-interference are located at the same wavelength, they experience the same phase shift as shown in Fig. 2(d). The self-interference can still be cancelled at the output of the BPD when the conditions in Eqs. (10) and (12) are met. Therefore, the optical fiber dispersion in uplink and downlink will not affect the SIC performance of the ROF system. It can be concluded that neither the fiber dispersion in the uplink nor the downlink will affect the SIC performance of the ROF system.

## 3. Experimental results

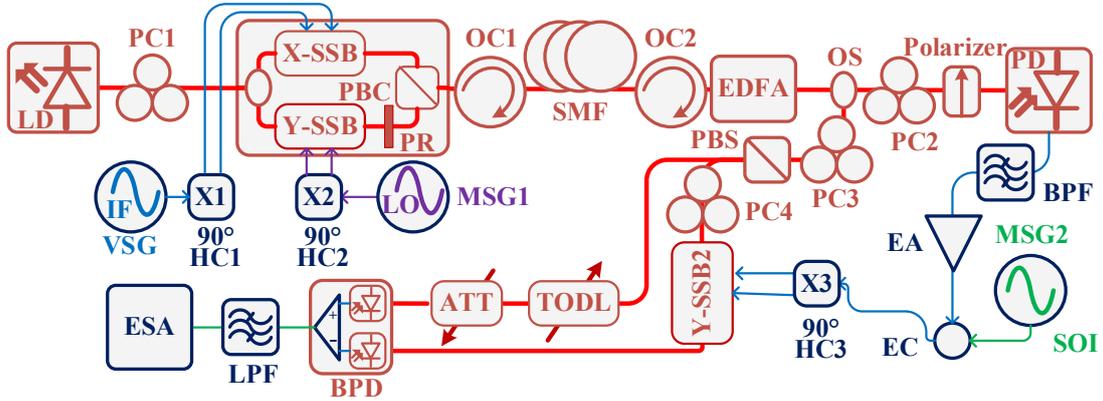

Fig. 3. Experimental setup of the proposed system. VSG, vector signal generator; MSG, microwave signal generator; BPF, bandpass filter; EA, electrical amplifier; EC, electrical coupler; LPF, lowpass filter; ESA, electrical spectrum analyzer.

An experiment based on the setup shown in Fig. 3 is performed. A CW light wave from an LD (ID Photonics CoBriteDX1-1-C-H01-FA) is injected into a DP-BPSK modulator (Fujitsu FTM7980EDA) via PC1. The downlink IF signal generated from a vector signal generator (VSG, Agilent N5182A) is sent to the two RF ports of the X-SSB modulator of the DP-BPSK modulator via 90° HC1 (Tamagawa electronics UPD-3025 1.7-2.3 GHz). The LO signal generated from a microwave signal generator (MSG1, Agilent 83630B) is applied to the two RF ports of the Y-SSB modulator of the DP-BPSK modulator via 90° HC2 (Macon Omni-Spectra FSC 16179 4-8 GHz). The optical signal from the DP-BPSK modulator is sent to the RU via a section of SMF. At the RU, the optical signal is amplified by an EDFA (Amonics AEDEF-PKT-DWDM-15-B-FA) and split into two paths via a 3-dB OS. One output from the OS is sent to a polarizer via PC2 and then applied to a PD (Discovery Semiconductors DSC-40S) to generate the downlink RF signal. The other output from the OS is demultiplexed into two orthogonally polarized branches (X and Y polarizations) by a PBS and PC3. The SSB-modulated optical signal in Y polarization is applied to a DD-MZM (Y-SSB2, Fujitsu FTM 7937EZ) as the optical carrier. The SOI is generated from MSG2 (HP 83752B), whereas the self-interference is simulated by the up-converted RF signal from the PD, which is filtered by a bandpass filter (BPF, 6.45-8.55 GHz) and amplified by an electrical amplifier (EA, CTT CLM/145-7039-293B, 5.85-14.85 GHz). The SOI and the self-interference are combined by an electrical coupler (EC, Narda 4456) and then applied to the Y-SSB2 modulator via 90° HC3 (Narda 4065 7.5-16 GHz). The output of the Y-SSB2 modulator is sent to one input port of the BPD (U2T BPRV2025), and the optical signal with its power attenuated and delayed by an ATT and a TODL in the X polarization is sent to the other input port of the BPD. The generated electrical signal from the BPD is then filtered by a lowpass filter (LPF, 3 GHz bandwidth), and then monitored by an electrical spectrum analyzer (ESA, Keysight N9020B). Due to the limited number of PBCs, PBSs, and PCs in our laboratory, the two optical signals in the two orthogonal polarization directions are not combined, transmitted back in the SMF, and split into two paths as shown in Fig. 1. Although the uplink transmission is omitted in the experiment, all the functions shown in Fig. 1 except the uplink transmission are all implemented in the experiment setup.

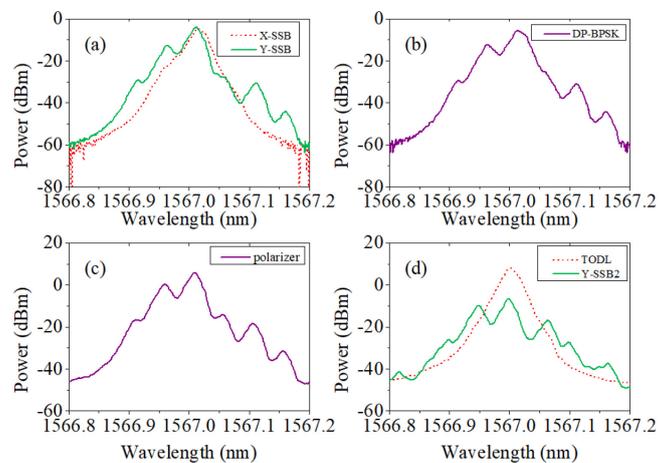

Fig. 4. Optical spectra of the optical signals from (a) the X-SSB modulator and the Y-SSB modulator; (b) the DP-BPSK modulator; (c) the polarizer; (d) the TODL and the Y-SSB2 modulator.

In the experiment, the wavelength and power of the CW light wave from the LD are set to 1567.0 nm and 10 dBm. The frequency and power of the IF signal are set to 2 GHz and 20 dBm, whereas those of the LO signal are set to and 6 GHz and 20 dBm. The corresponding optical spectra at the X-SSB modulator and the Y-SSB modulator observed by the optical spectrum analyzer (OSA, ANDO AQ6317B) with a resolution of 0.01 nm is shown in Fig. 4(a). The optical signal in the X polarization is modulated by the IF signal, which contains the optical carrier and the -1st-order optical sideband as shown by the red dash line. The green solid line shows the optical signal in the

Y polarization modulated by the LO signal, which contains the optical carrier and the +1st-order optical sideband. Due to the limited resolution of the OSA, the optical carrier and the optical sideband cannot be clearly distinguished in the X polarization because the wavelength spacing between then is only 2 GHz. Fig. 4(b) shows the spectrum of the polarization multiplexed optical signal at the output of the DP-BPSK modulator. The optical signal is then split into two parts. The part for the generation of the up-converted downlink RF signal is combined at the polarizer, with its optical spectrum shown in Fig. 4(c). The other part of optical signal is demultiplexed into two orthogonally polarized branches, which are respectively used as the reference optical signal for self-interference cancellation and the optical carrier for the uplink modulation and frequency down-conversion. The reference optical signal in the X polarization are adjusted by the TODL and the ATT to match the delay and amplitude of the self-interference. In our experiment, since a narrowband self-interference is employed, no RF phase shifter is used. Therefore, the TODL should be adjusted to compensate not only the time delay but also the phase shift according to Eq. (11). The optical spectrum after the TODL is shown by the red dash line in Fig. 4(d). The optical signal in Y polarization is SSB-modulated by the received signal and generate two optical sidebands at frequencies of $f_c$-$f_{IF}$ and $f_c$-$f_s$, which is shown by the green solid line in Fig. 4(d).

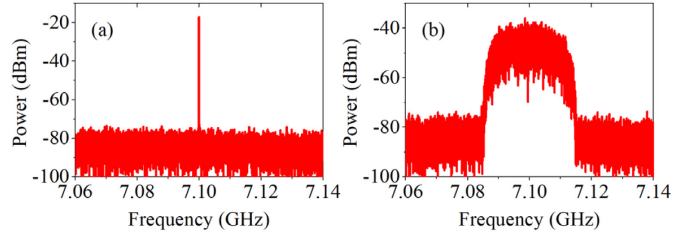

Fig. 5. Electrical spectra of the downlink RF signals at the output of the BPF when the IF signal is (a) a single-tone signal; (b) a 20-MBaud 16-QAM signal.

Then, the downlink signal generation is experimentally verified. Due to the limitation of the operating bandwidth of the 90° HCs, the BPF, and the LPF that are available in our laboratory, the frequencies of the IF signal, the LO signal, and the RF signal must be carefully selected so as to verify the experimental setup in Fig. 3. In this experiment, the central frequencies of the IF signal and the LO signal are set to 2.1 GHz and 5 GHz, respectively, and the power of the two signals is set to 10 dBm. The generated up-converted downlink RF signals at the output of the BPF are shown in Fig. 5. When a single-tone signal is employed as the IF signal, a single-tone RF signal with a frequency at 7.1 GHz can be observed as shown in Fig. 5(a). When the IF signal is replaced by a 20-MBaud 16-QAM signal, the generated downlink RF signal that is centered at 7.1 GHz is shown in Fig. 5(b). It can be seen that the power of the up-converted RF signal from the BPF is very low, so it is then amplified by the EA with 39-dB gain which is then used as the self-interference in the experiment.

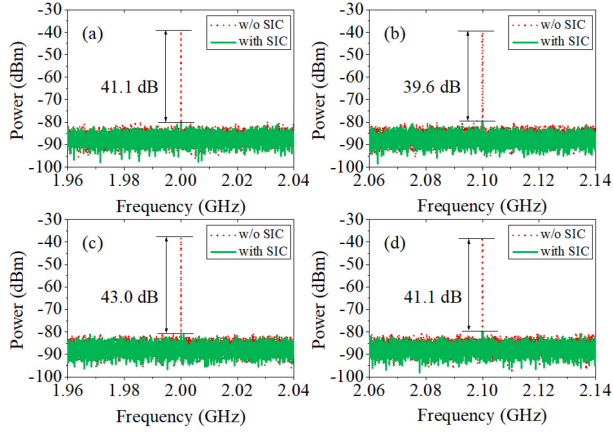

Fig. 6. Electrical spectra of the down-converted IF signal in the uplink with and without SIC when the frequencies of the IF signal and the LO signal are (a) 2 GHz, 5 GHz; (b) 2.1 GHz, 5GHz; (c) 2 GHz, 6 GHz; (d) 2.1 GHz, 6 GHz. No SMF is inserted.

The ability of SIC and frequency down-conversion in the uplink is also demonstrated. A 6-dBm single-tone signal centered at 2 GHz is used as the IF signal and it is up-converted to 7 GHz by an LO signal with a frequency of 5 GHz and a power of 17.3 dBm. The 7-GHz signal at the output of the EA is used as the self-interference and applied to Y-SSB2 via 90° HC3. The SOI is not considered in this case. Under these circumstances, the electrical spectrum at the output of the BPD without SIC is shown in the red dash line in Fig. 6(a), where the self-interference at 2 GHz after the frequency down-conversion is observed. When the SIC is enabled, the electrical spectrum at the output of the BPD is shown in the green solid line in Fig. 6(a). It can be seen that the self-interference is well canceled with a cancellation depth of more than 40 dB. Furthermore, the frequency tunability of the system is also verified by changing the frequencies of the IF signal and the LO signal. Figure 6(b) to (d) show the electrical spectra at the output of the BPD with and without SIC under different situations. The frequency of the IF signal is 2 or 2.1 GHz and the frequency of the LO is 5 or 6 GHz. It can be seen that the cancellation depths are all around 40 dB.

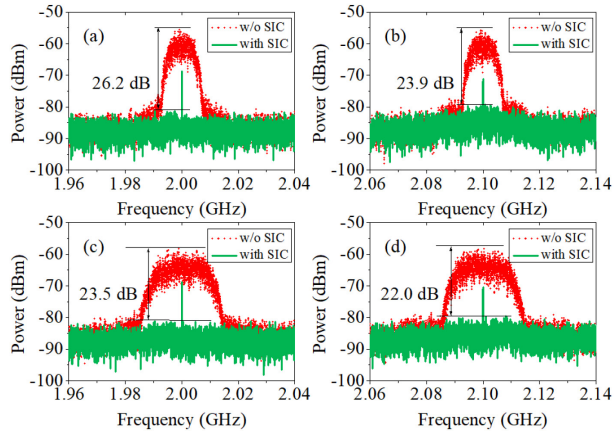

Fig. 7. Electrical spectra of the down-converted IF signals in the uplink with and without SIC when the IF symbol rate, the central frequencies of the IF signal and the LO signal are (a) 10-MBaud, 2 GHz, 6 GHz; (b) 10-MBaud, 2.1 GHz, 6 GHz; (c) 20-MBaud, 2 GHz, 6 GHz; (d) 20-MBaud, 2.1 GHz, 6GHz. No SMF is inserted.

Then, the SIC performance of the system is further verified by employing vector signals as the self-interference. First, a 16-QAM signal centered at 2 GHz with a symbol rate of 10 MBaud and a power of 6 dBm is used as the input IF signal for the downlink. The frequency and power of the LO signal is set to 6 GHz and 17.3 dBm. A -22-dBm single-tone sinusoidal signal at 2 GHz is employed as the SOI in this case. The electrical spectrum at the output of the BPD with and without SIC is shown in Fig. 7 (a). It can be seen that when the self-interference is enabled, the SOI is completely submerged by the self-interference. After SIC, the self-interference is suppressed by more than 26 dB, and the SOI can be clearly observed. Then, the central frequency of the IF signal is changed to 2.1 GHz, and a cancellation depth of 23.9 dB is obtained as shown in Fig. 7(b). The SIC performance of the system is further studied by increasing the symbol rate of the self-interference to 20 MBaud. As shown in Figs. 7(c) and (d), the cancellation depths are about 23.5 dB or 22 dB, respectively, when the central frequency of self-interference is 2 GHz or 2.1 GHz. Compared with the results in Figs. 7(a) and (b), it can be seen that the cancellation depth slightly deteriorates when the symbol rate of the self-interference increases. The reason is that a TODL is used in the experiment to compensate the -5π/4 phase shift in the system instead of a phase shifter. As the bandwidth increases, the phase compensation by the TOLD will be more inaccurate, which leads to the decrease of the cancellation depth.

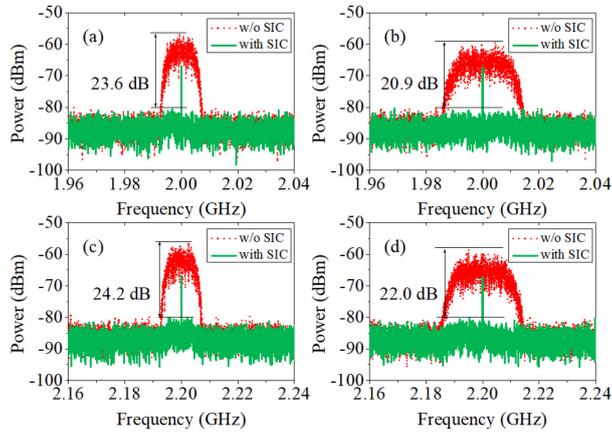

Fig. 8. Electrical spectra of the down-converted IF signals in the uplink with and without SIC when the IF symbol rate, the central frequencies of the IF signal and the LO signal are (a) 10-MBaud, 2 GHz, 6 GHz; (b) 20-MBaud, 2 GHz, 6 GHz; (c) 10-MBaud, 2.2 GHz, 6 GHz; (d) 20-MBaud, 2.2 GHz, 6GHz. A section of 4.1-km SMF is inserted.

Finally, a section of 4.1-km SMF is inserted in the system to further verify the SIC performance. In the experiment, the frequency of the LO signal is fixed at 6 GHz while the frequency and the symbol rate of the IF signal are changed. The electrical spectra at the output of the BPD are shown in Fig. 8. When the central frequency of the IF signal is set to 2 GHz and the symbol rates of the input IF signals are 10 MBaud and 20 MBaud, the cancellation depths are 23.6 dB and 20.9 dB, respectively. Then, the central frequency of the IF signal is adjusted to 2.1 GHz, the cancellation depths are 24.2 dB and 22 dB. Compared with the results without fiber transmission, no significant performance degradation is observed.

In the experiment, the operating bandwidths of the three 90° HCs need to have a certain relationship. Due to the limitation of the such electrical devices available in our laboratory, the system is only demonstrated by employing an IF signal at around 2 GHz, and an RF signal around 6 to 8 GHz. In fact, these frequencies can be adjusted in a wider range if three 90° HCs with better bandwidth matching are employed. As discussed in Eq. (11), a -5π/4 phase shift needs to be compensated so as to make the system suitable for wideband systems. Because the electrical phase shifter in our laboratory is all based on time delay, we only demonstrate the SIC performance of a narrowband system.

## 4. Conclusions

We have proposed and experimentally demonstrated a photonic approach for SIC incorporated in an IBFD ROF system. All the functions, including overcoming the influence on the SIC and the distortion of the SOI caused by the fiber dispersion, realizing SIC in conjunction with frequency conversion, and moving the SIC from the RU to CO to simplify the RU, are simultaneously realized in the system, which provides a feasible solution for the SIC in IBFD ROF systems. An experiment is carried out. The SIC performance of the proposed approach is studied by employing a single-tone signal and a vector signal as the self-interference. The cancellation depth is around 40 dB for the single-tone self-interference and around 22 dB for the 20-MBaud 16-QAM modulated self-interference.

**Funding**

This work was supported by the Natural Science Foundation of Shanghai [grant number 20ZR1416100]; the National Natural Science Foundation of China [grant number 61971193]; the Open Fund of State Key Laboratory of Advanced Optical Communication Systems and Networks, Peking University, China [grant number 2020GZKF005]; and the Fundamental Research Funds for the Central Universities.

**Conflicts of interest**

The authors declare no conflicts of interest.